# SECURE AND RELIABLE ROUTING IN MOBILE AD-HOC NETWORKS

Rachid Haboub and Mohammed Ouzzif

RITM laboratory, Computer science and Networks team.
ENSEM - ESTC - UH2C, Casablanca, Morocco.

rachidhaboub@hotmail.com and ouzzif@yahoo.com

## ABSTRACT

*The growing diffusion of wireless-enabled portable devices and the recent advances in Mobile Ad-hoc NETworks (MANETs) open new scenarios where users can benefit from anywhere and at any time for impromptu collaboration. However, energy constrained nodes, low channel bandwidth, node mobility, high channel error rates, channel variability and packet loss are some of the limitations of MANETs. MANETs presents also security challenges. These networks are prone to malicious users attack, because any device within the frequency range can get access to the MANET. There is a need for security mechanisms aware of these challenges. Thus, this work aims to provide a secure MANET by changing the frequency of data transmission. This security approach was tested, and the results shows an interesting decreased of throughput from malicious node when the number of frequency used is increased, that way the MANET will not waste it's resources treating malicious packets. The other contribution of this work is a mobility aware routing approach, which aims to provide a more reliable routing by handling effectively the nodes mobility.*

## 1. INTRODUCTION

Worldwide sales of smart phones, laptops, and PDAs have increased exponentially each year since their introduction. These mobile devices can be used to form MANETs. A MANET consists of arbitrary deployed communicational devices, such as cellular phones, Personal Digital Assistants (PDAs), laptop, etc. it is a multi-hop wireless network where all nodes cooperatively maintain the network connectivity. The mobile nodes are capable of connecting and communicating with one another using limited-bandwidth radio links. These types of networks are useful in any situation where temporary network connectivity is needed and in areas with no prefixed infrastructure, such as disaster relief where existing infrastructure is damaged, or military applications where a tactical network is required.

In the battlefield, typically in a foreign land, one may not rely on the existing infrastructure. In these situations, establishing infrastructure is not practical in terms of expenditure and time consumed. Hence, providing the needed connectivity and network services becomes a real challenge. In a wireless ad hoc network where pairs of mobiles communicate by exchanging a variable number of data packets along routes set up by a routing algorithm, reliability may be defined as the ability to provide high delivery rate, that is, to deliver most of the data packets in spite of faults breaking the routes or buffer overflows caused by overloaded nodes. Given the intrinsic nature of wireless ad hoc networks, reliability is a major issue. Figure one gives an overview of MANETs applications.

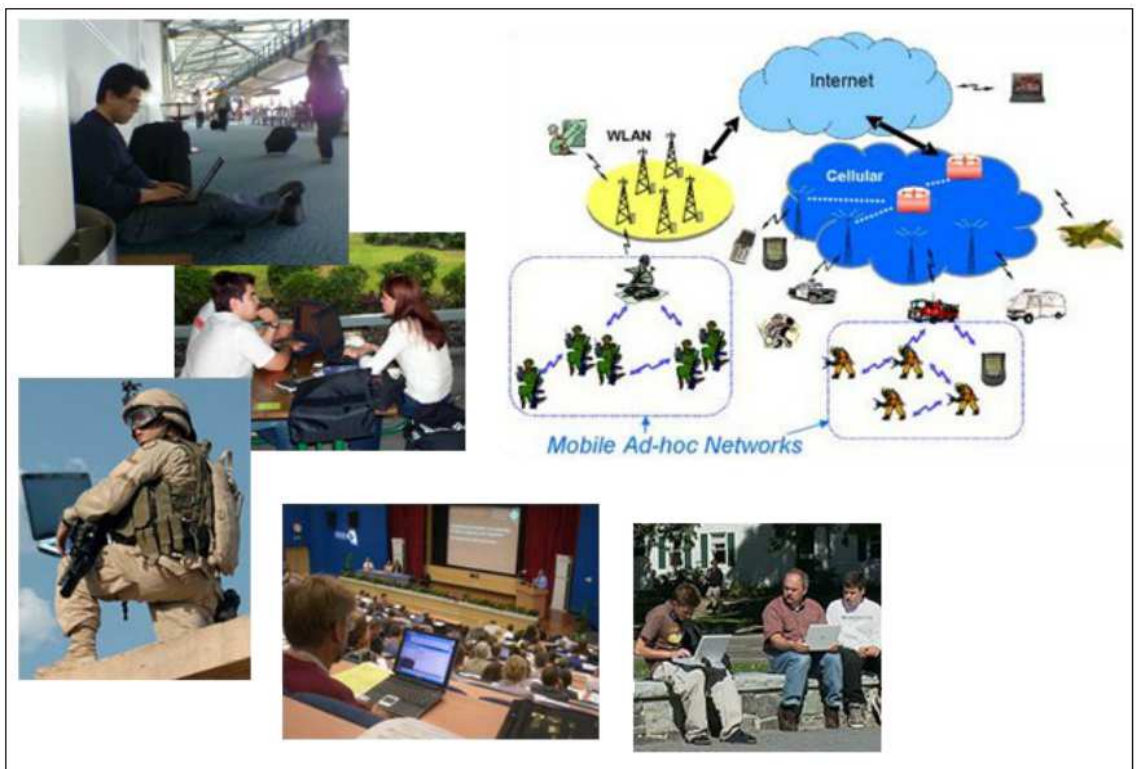

Figure 1. MANETS applications

Links failures may occur due to interferences on the wireless medium, or, most probably, to nodes mobility, when pairs of nodes move out of the reciprocal transmission range or are shadowed by obstacles. MANETs do not only provide dynamic infrastructure networks but also allow the flexibility of wireless devices mobility. MANETs differ significantly from existing networks. First, the topology of the nodes in the network is dynamic. Second, these networks are self-configuring in nature and do not require any centralized control or administration. Such networks do not assume all the nodes to be in direct transmission range of each other. Hence these networks require specialized routing protocols that provide self-starting behaviour.

However energy constrained nodes, low channel bandwidth, node mobility, high channel error rates, and channel variability are some of the limitations of MANETs. Under these conditions, existing wired network protocols would fail or perform poorly. Thus, MANETs require specialized routing protocols. In the battle field, where there is no pre-existing infrastructure, security of all the MANET's components is primordial. For example, the enemy may attack a set of mobile devices (for malicious purposes, in order to get access to a specific area, etc.). Existing security approaches uses mainly encryption keys, which need too much resources. And as we know, MANET's resources are limited [13].

 In this paper we use a security and a security and mobility aware approach, to improve the AODV (Ad-hoc On-demand Distance Vector) [10] routing protocol, in order to be protected from the resource exhaustion attack, by periodically changing the packet transmission frequency, and to have a more reliable network. This paper is organized as follow: section two present the problematic, section three gives some related works. Section four present the proposed approach, section five gives a simulation of the security approach. Finally the last section gives a conclusion and future works.

## 2. PROBLEMATIC

Now a day, MANETs are used in many critical fields such as emergency, military, etc. In such networks, security is very important, in order to be protected from malicious nodes attacks. There are many kinds of attacks: node outage, link layer jamming, collision attack, traffic manipulation and resource exhaustion [20].

Node outage consists of stopping the functionality of the MANET's components, such as a cluster-leader, by physically or logically damaging the network. Link layer jamming consist of finding data packet and to jam it [7], this kind of attack can cause colliding packets during transmission, exhausting nodes' resources and confusions.

Collision attacks consist of creating interferences in the network, by changing packet's fields and altering the "Ack message". This may causes a corruption, a cripple and discarding the transmitted packets. It can also cost an energy exhaustion which is cost effective [9].

Traffic manipulations consist of regular monitoring transmissions and computing some parameters based on affected MAC protocol carefully, by doing a time adjustment, in order to transmit messages just at the moment when normal nodes do so. This decrease the signal quality, network availability and MANET's performance, it also breaks the protocols operations, destroy the traffic and create confusions. This kind of attack is similar to the collision attack.

Resource exhaustion attack (in which we are interested) consists of doing repeated collisions and continuous retransmission out-of-date, dead and corrupted packets until the sensor node death. We consider as a scenario of MANET of 24 nodes and a malicious one (node 25). If the malicious node wants to do a resource exhaustion attack on node 0, it will at first search how to reach it. In order to send malicious packets, the malicious node begins by broadcasting a RREQ (Route REQuest) message, to find a path leading to node 0 (figure 2).

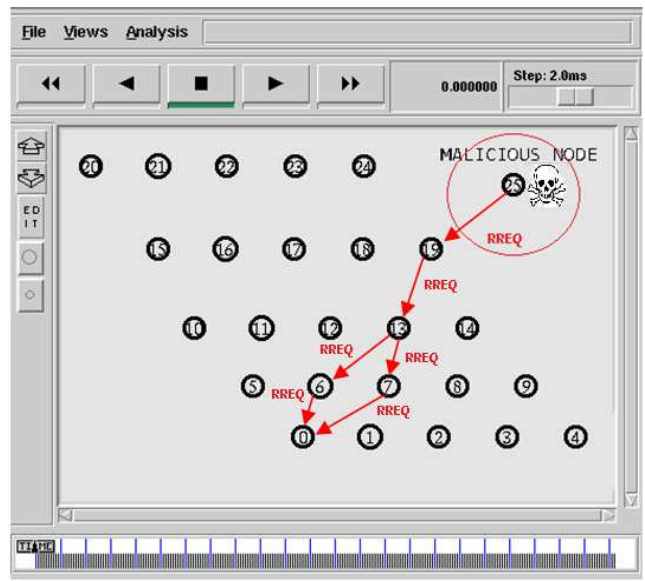

Figure 2. Malicious node sending Route REQuests to get access to node 0.

Next, a RREP (Route REPly) message will be received by the malicious node with the destination paths to node 0 (figure 3).

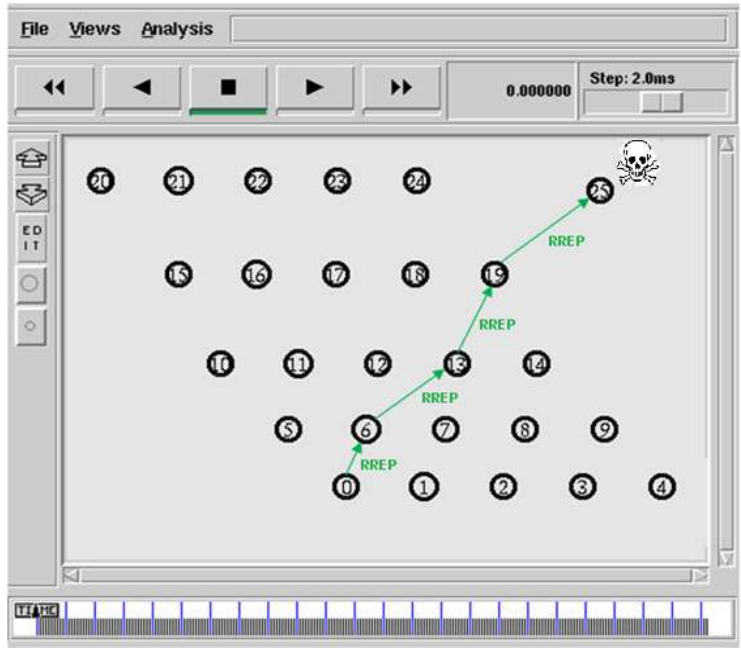

Figure 3.  RREP message from destination to malicious node.

Once node 0 is localized and the path to reach it is defined, the malicious node will be able to send his malicious packets to node 0 (figure 4).

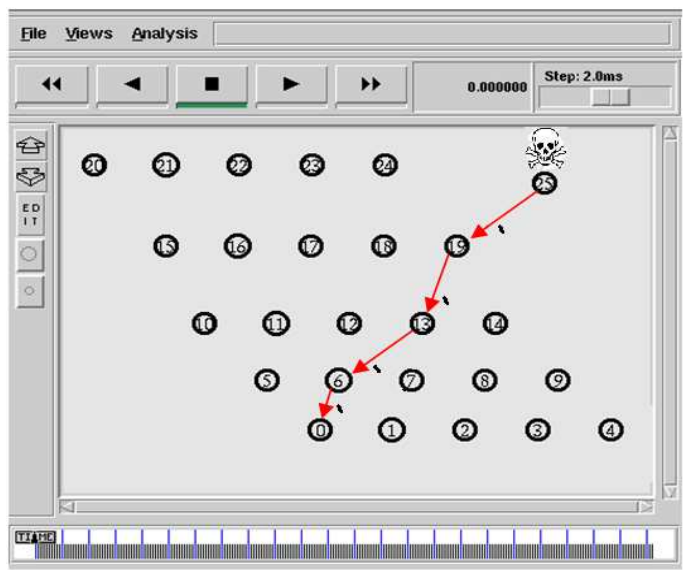

Figure 4.  Malicious node trying to do a resource exhaustion attack to node 0.

## 3. RELATED WORKS

There are a lot of attacks types described in [5]. We are interested in the "Resource Exhaustion" attack. Existing security approaches require a certain amount of resources, including data memory, energy, etc. However, currently these resources are very limited in a tiny wireless sensor. The most common security mechanism is encryption techniques [1] which require using security keys and encrypting data, which consume the memory storage space inside the device.

Wood and Stankovic [3] define a kind of denial of service attack as "any event that diminishes or eliminates a network's capacity to perform its expected function", denial of service attacks are not a new techniques, although this is still an open problem to the network security community. Unfortunately, wireless sensor networks cannot afford the computational overhead necessary in implementing many of the typical defensive strategies. LEAP [15] establishes different security requirements for each type of message, therefore, it uses four types of keys for each sensor node, and nodes need more storage capabilities, each sensor node has to store four types of keys, and it needs efficient mechanisms to update the keys.

## 4. APPROACH

### 4.1. The security approach

There are two main categories of routing protocol in MANET (figure 5): proactive and reactive [2]. Proactive protocols maintain fresh lists of destinations and their routes by periodically distributing routing tables throughout the network. Reactive protocols find a route on demand (only when needed) by flooding the network with Route Request packets. We have chosen the second kind of protocols which is suitable for the limited resources of MANET's. We adopt AODV which do not maintain routing information, but depend totally on needs to communicate.

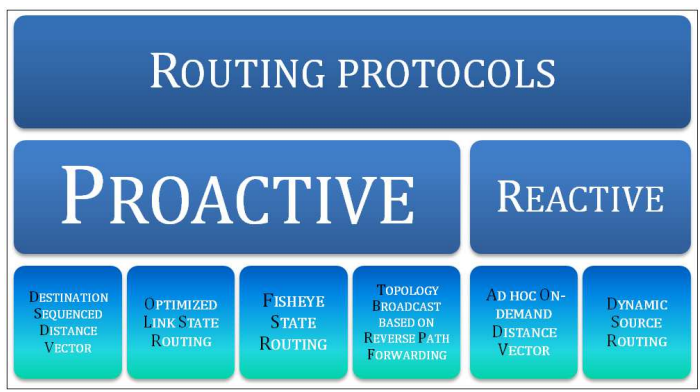

Figure 5. Routing protocols classification

AODV is suitable for the limited resources of MANET. It does not maintain any routing information but totally depends on needs to communicate with its neighbourhood. AODV broadcast discovery packets only when necessary. It is a simplest and widely used reactive routing protocol, which consume low energy, because it searches for a path to reach the destination on demand (only when a node needs to send packets). AODV do not store periodically the paths to destinations in routing tables.

In order to avoid the resource exhaustion attack, we propose to use different frequencies. This technique was suggested in the RFID (Radio Frequency IDentification) systems [6]. We adopt this mechanism in the AODV routing protocol, in order to ensure a secure AODV (SAODV). It is an efficient way to provide security in MANET's. If we use for example 30 frequencies and if intruders get access to the channel, it will only affect a particular channel, and there will be 29 channels still available for data transmission. If the number of frequency used are increased or/and time set of each frequency is random, the probability of intruder accessing the channels will be small.

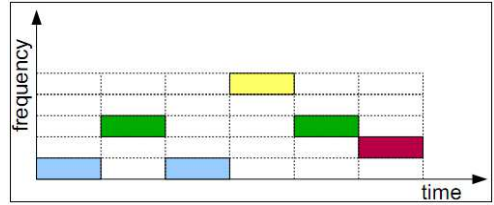

Figure 6.  Frequency changing over time

AODV packet format consist of headers and data.  There  are  many  types  of  header available such as common header, IP header, TCP header, RTP header and Trace header. One can add his own header too by creating a new header in the NS2 (Network Simulator). The type  of header  used  in this  project  is  common  header.  Common header uses many fields to store packets information.

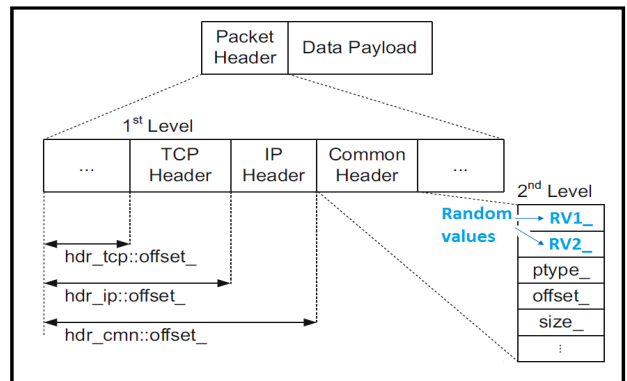

Figure 7.  Packet format

In order to implement the security approach, we suggest adding two fields in the common header part of the packet header. This fields store two random values RV1 and RV2 (figure 7), generated before the packet transmission. Those random values allow the receiver to know malicious packets frequency in order to reject them. That way, the receiver will not waste his resources in treating malicious packets.

```
// Creating  two variables to store the random number:
        Double  RV1;
        Double  RV2;

// Setting two random values (the random values are between 0 and 1):
        RV1 = random number;
        RV2 = random number;

// Setting the frequency according to r:
        if (RV1 <= RV2)
        { Send packets using frequency number 1; }
        else
        { Send packets using frequency number 2; }
```

Figure 8.  Changing frequency before the packet transmission

We have added in the AODV function which is responsible for packets transmission, the algorithm described in figure 7. This algorithm consists of generating two random values and sending packets in a specific frequency according to them. Before transmitting a packet, SAODV (Secure AODV) transmission function will set the frequency to either 1 or 2. The variables RV1 and RV2 which has double data type, will keep the random numbers generated, in order to use them at the reception side.

```
// Verification of the random values:
If (RV1 + RV2 <= 2) && (RV1 + RV2 >= 0)
{
            // Setting the frequency according to the random number r:
            If (RV1 <= RV2) { Random_Frequency =1;}
            else              { Random_Frequency =2;}
}

Else { Drop (packet); }

// Checking if the Random_Frequency is the same as the incoming packet
frequency:
If ( Random_Frequency != Incoming packet frequency )
            { Drop (packet); }

Else // Continue with AODV.
```

Figure 9.  Frequency verification at the packet reception

When a packet is received, SAODV receiving function will check the random values and the incoming frequency (figure 9). If the packet frequency is not the same as the frequency according to the random number, then the packet will be dropped. If the frequency is the same to the frequency generated by the random numbers, the packet will be accepted.

## 4.2. The mobility aware approach

If the motion parameters of two neighbouring nodes like speed, direction, radio propagation range are known, the duration of time these two nodes will remain connected can be determined. Assume two nodes i and j within the transmission range of each other. Let (xi, yi) be the coordinates of node i and (xj, yj) be the coordinates of node j. Let Vi and Vj be the speeds, ɵi (0 ≤ ɵi ) and ɵj (ɵj ≤ 2∏) be the directions of motion for nodes i and j, respectively. The amount of time two mobile hosts will stay connected, is predicted by the formula given by equation:

$$LET = \frac{-(ab+cd) + \sqrt{(a^2+c^2)r^2 - (ad-bc)^2}}{a^2+c^2}$$

$$a = V_r \cos\theta - V_s \cos\theta$$
$$b = X_r - X_s$$
$$c = V_r \sin\theta - V_s \sin\theta = V_{Y_r} - V_{Y_s}$$
$$d = Y_r - Y_s$$

Figure 10.  Link expiration time formula

r is the transmission range of a wireless node with an Omni-directional antenna, which is 250 m. Vs and Vr are the velocities of the sender and receiver respectively. Ѳ is the direction of motion of nodes. (Xs,Ys) and (Xr,Yr) are the coordinates of the sender and receiver respectively.

Parameter "a" is the relative velocity of the receiver node with respect to the sender node along Y axis. "b" is the parameter used to determine the distance of the receiver node from the sender node along X axis. The third parameter used to determine LET is "c", which is the relative velocity of receiver node with respect to the sender node along Y axis. "d" is the distance of the receiver node from the sender node along Y axis.

Consider a MANET consisting of four nodes illustrated in figure eleven, figures a, b and c represent the network topologies for a non mobility aware routing at times t, t+1 and t+2, respectively. Figures d, e and f narrate the expected network topologies for the proposed mobility aware routing approach at times t, t+1and t+2, respectively. Node 0 and Node 3 are assumed to be sender and receiver nodes respectively. A non-mobility aware protocol would use route 0-1-2. If node 1 is moving away out of the transmission range of node 3, the link 1-3 breaks. This event initiates route maintenance activity which results in heavy control traffic generated by node 0 and node 3 in an attempt to revive the broken link but in vain. It forms route 0-2-3 to retain the network data transmission.

Apart from high control traffic generated, the active transmission of data through these links during a link disconnection results in loss of data packets. Both the excess control overhead generated to revive the broken links and the data packet loss could have been avoided if a more reliable route 5-4-3-1 was formed instead of 5-4-2-1. This can be achieved with the implementation of the mobility aware routing algorithm in the underlying routing protocol. That way, the fast moving node, node 2, is eliminated from route discovery process by node 1, and the routing protocol forms the route through node 3 instead. The proposed mobility aware approach drop packets when node mobility does not permit to form a link for the necessary amount of time.

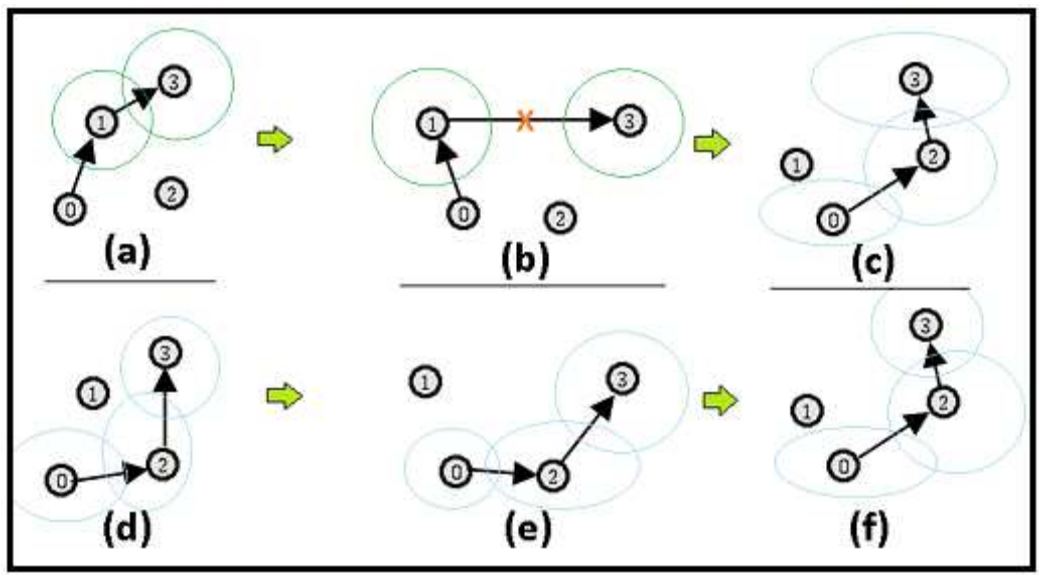

Figure 11. The mobility aware routing approach

This link survives through the data transmission and thus, eliminates the control overhead generated by the non mobility aware protocol, to resuscitate a link breakage. Thus, the mobility aware approach promises to restrict the number of unreliable routes based on node mobility.

The decision-making parameter for the proposed approach is the route reliability. In this research, route reliability is measured by the amount of time two nodes can be connected without a link disconnection. The link connectivity is determined using the link expiration time.

```
//Getting the receiver coordinates
Xr, Yr, Zr

//Getting the receiver velocities
V(Xr), V(Yr), V(Zr)

//Getting the sender coordinates
Xs, Ys, Zs

//Getting the sender velocities
V(Xs), V(Ys), V(Zs)

//Calculating the Link expiration time
double a = V(Xr)-V(Xs);
double b = Xr-Xs;
double c = V(Yr)-V(Ys);
double d = Yr-Ys;

// The average transmission range of a wireless node with an Omni-directional antenna is 250 m.
double r = 250;

double P = (((a*a)+(c*c))*(r*r))-(((a*d)-(b*c))*((a*d)-(b*c)));
float Q;

if (P>=0) {Q = sqrt(P);        }
else        { Q = sqrt(-(P));        }

if (((a*a)+(c*c)) == 0.0) {// LET will have an infinite value}
else { LET = (-1*((a*b)+(c*d))+Q)/((a*a)+(c*c));}

//If LET is too low, drop the packet
```

Figure 12. The mobility aware approach algorithm

## 5. SIMULATION

We use the NS-2 simulator [11]. NS-2 takes as an input a TCL file (in which we implement the scenario). NS2 consists of two key languages: C++ and Object-oriented Tool Command Language (OTcl). While the C++ defines the internal mechanism of the simulation objects, the OTcl sets up simulation by assembling and configuring the objects as well as scheduling discrete events. After simulation, NS2 outputs a trace file, which can be interpreted by many tools, such as NAM and Xgraph. We create a simulation scenario using NS-2 Scenario Generator [12].

Table 1 shows the network parameter definition in the TCL file. The first parameter tells the simulator that nodes transmits and receives packets through wireless channels. We have used the IEEE 802.15.4 standard, which specifies the media access control and the physical layer. val(nn) is the number of nodes, which is set to 25. val(rp) is set to the SAODV protocol, which represent the routing protocol used in the simulation. val(x) and val(y) are equal to 50 meter. So 50 m² is the simulation area. val(stop) represent the simulation time, and is equal to 50 second.

Table 1.  Network parameter definition.

| Parameter | Suggested Value | Description |
|---|---|---|
| val (chan) | Channel/Wireless Channel | Channel type |
| val (mac) | Mac/802_15_4 | IEEE standard |
| val (nn) | 25 | Number of nodes |
| val (rp) | SAODV | Routing protocol |
| val (x) | 50 | Setup topography object |
| val (y) | 50 | Setup topography object |
| val (stop) | 50 | Simulation time |

Figure 11 shows the NS-2 trace file format. The first field is event, it gives many possible symbols ( 'r', 'd', etc. ). These symbols may correspond for example to received and dropped packets. The second field gives the time at which the event occurs. The third field gives the source node at which the event occurs. The fourth field gives the destination node at which the event occurs. The fifth field shows information about the packet type, whether it's a UDP or a TCP packet. The sixth field gives the packet size. The seventh field gives information about some flags. The Fid field is the flow Id, it can be used for specifying the colour of flow in NAM display. The ninth field is the source address. The tenth field is the destination address. The eleventh field is the network layer protocol's packet sequence number, and the last field shows the unique id of a packet.

| Event | Time | Source | Destination | Pkt type | Pkt size | Flags | Fid | Src addr | Dst addr | Seq num | Pkt id |
|---|---|---|---|---|---|---|---|---|---|---|---|

Figure 13.  NS-2 Trace File format

We use the Java-Trace-Analyzer to interpret the trace file generated by the simulation. The simulation results presented in figure 12 shows that the throughput of dropping packets (coming from the malicious node) at the destination node is low using the AODV protocol. The throughput of dropping packets at the receiving node becomes high using the security approach. This means that using this approach allow rejecting malicious packets. Then malicious node will not be able to do resource exhaustion on the MANET.

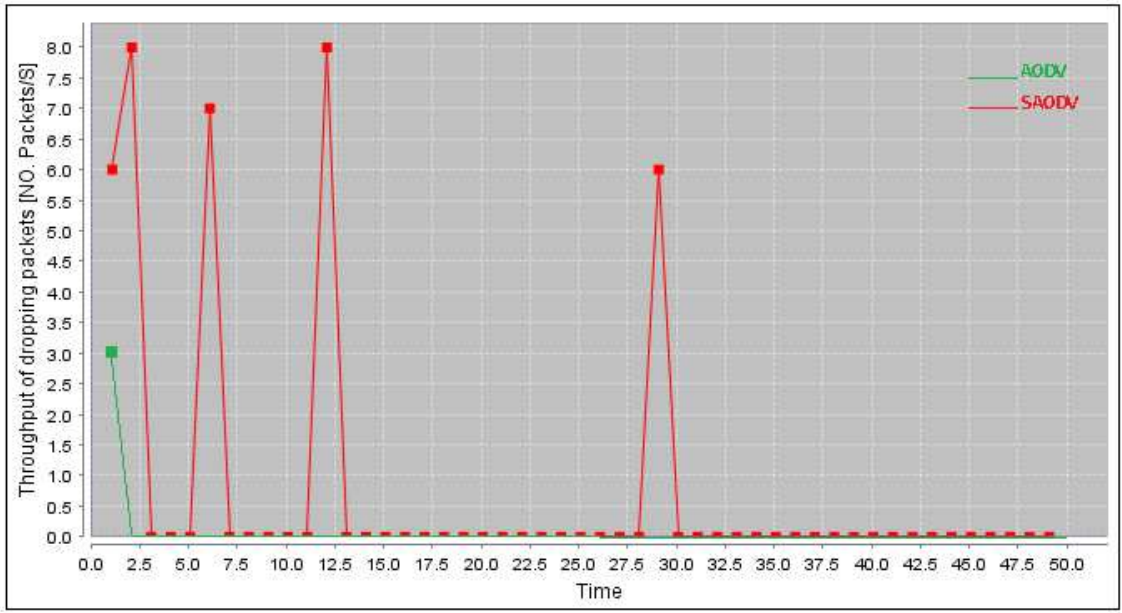

Figure 14.  Throughput of dropping packets (coming from the malicious node) at the receiving node

The simulation results presented in figure 13 shows that using SAODV, the victim node will not waste its energy. However, when we use AODV, the victim node wastes all its energy in 60 minutes.

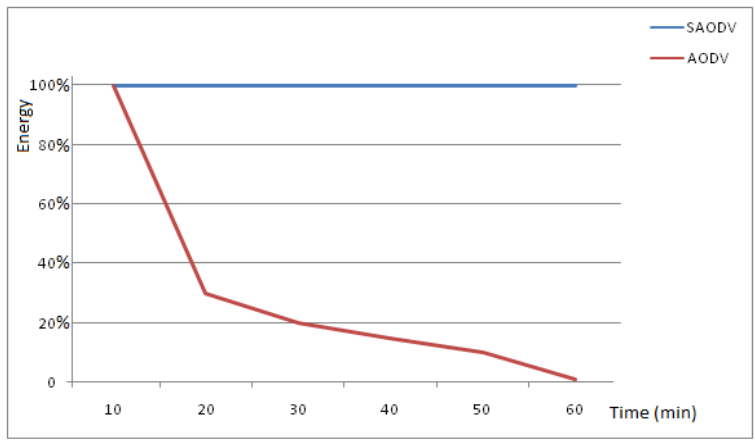

Figure 15.  Energy level over time of the victim node

Figure 16 shows that using the proposed approach, the packets lost decrease.

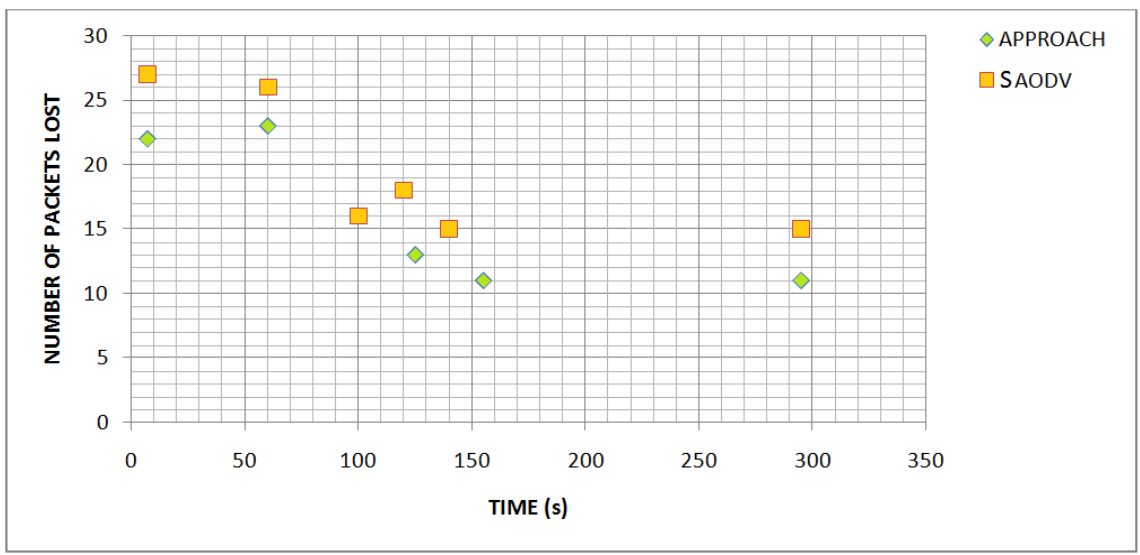

Figure 16.  Packet loss versus time

## 6. CONCLUSION AND FUTURE WORKS

In this paper, we have used an approach based on changing the packet transmission frequency and mobility awareness in the AODV protocol, which is a reactive protocol. The goal is to avoid the resource exhaustion attack and improve the routing process. We have shown a resource exhaustion attack scenario, which we have simulated using NS-2. Simulation results show that a victim node will not waste its resources treating malicious packets. Changing the frequency for transmission packets in order to secure a MANET from the resource exhaustion attack, may be useful for many fields, such as in the military. For future work we intend experimenting the changing frequency approach combined with the proposed mobility aware approach, using more metrics and criteria (such as nodes energy, mobility, connectivity, vicinity, etc.).

**Authors**

**Rachid Haboub** is a full time Ph.D student. He received the Master degree in computer science, from Hassan II University, Ben M'sik faculty of Morocco in 2009. His research spans wireless communication.

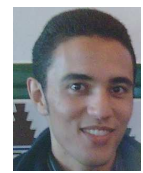

**Dr. Mohammed Ouzzif** is a professor in the computer science department of the higher school of technology of Casablanca - Hassan II university of Morocco.

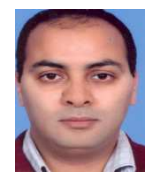